

%
%
\input harvmac
%
%
%
%
%
%
%
%
%
\newif\ifdraft

\noblackbox
\catcode`\@=11
\newif\iffrontpage
%
\ifx\answ\bigans
\def\titleft{\titsm}
\magnification=1200\baselineskip=15pt plus 2pt minus 1pt
%
\advance\hoffset by-0.075truein
\hsize=6.15truein\vsize=600.truept\hsbody=\hsize\hstitle=\hsize
\else\let\lr=L
\def\titleft{\titla}
\magnification=1000\baselineskip=14pt plus 2pt minus 1pt
%
\vsize=6.5truein
\hstitle=8truein\hsbody=4.75truein
\fullhsize=10truein\hsize=\hsbody
\fi
\parskip=4pt plus 10pt minus 4pt

\font\titla=cmr10 scaled\magstep3
\font\tenmss=cmss10
\font\absmss=cmss10 scaled\magstep1
\newfam\mssfam
\font\footrm=cmr8  \font\footrms=cmr5
\font\footrmss=cmr5   \font\footi=cmmi8
\font\footis=cmmi5   \font\footiss=cmmi5
\font\footsy=cmsy8   \font\footsys=cmsy5
\font\footsyss=cmsy5   \font\footbf=cmbx8
\font\footmss=cmss8
\def\footfont{\def\rm{\fam0\footrm}
\textfont0=\footrm \scriptfont0=\footrms
\scriptscriptfont0=\footrmss
\textfont1=\footi \scriptfont1=\footis
\scriptscriptfont1=\footiss
\textfont2=\footsy \scriptfont2=\footsys
\scriptscriptfont2=\footsyss
\textfont\itfam=\footi \def\it{\fam\itfam\footi}
\textfont\mssfam=\footmss \def\mss{\fam\mssfam\footmss}
\textfont\bffam=\footbf \def\bf{\fam\bffam\footbf} \rm}
\def\tenpoint{\def\rm{\fam0\tenrm}
\textfont0=\tenrm \scriptfont0=\sevenrm
\scriptscriptfont0=\fiverm
\textfont1=\teni  \scriptfont1=\seveni
\scriptscriptfont1=\fivei
\textfont2=\tensy \scriptfont2=\sevensy
\scriptscriptfont2=\fivesy
\textfont\itfam=\tenit \def\it{\fam\itfam\tenit}
\textfont\mssfam=\tenmss \def\mss{\fam\mssfam\tenmss}
\textfont\bffam=\tenbf \def\bf{\fam\bffam\tenbf} \rm}
\ifx\answ\bigans\def\abstractfont{\tenpoint}\else
\def\abstractfont{\def\rm{\fam0\absrm}
\textfont0=\absrm \scriptfont0=\absrms
\scriptscriptfont0=\absrmss
\textfont1=\absi \scriptfont1=\absis
\scriptscriptfont1=\absiss
\textfont2=\abssy \scriptfont2=\abssys
\scriptscriptfont2=\abssyss
\textfont\itfam=\bigit \def\it{\fam\itfam\bigit}
\textfont\mssfam=\absmss \def\mss{\fam\mssfam\absmss}
\textfont\bffam=\absbf \def\bf{\fam\bffam\absbf}\rm}\fi
%
\def\f@@t{\baselineskip10pt\lineskip0pt\lineskiplimit0pt
\bgroup\aftergroup\@foot\let\next}
\setbox\strutbox=\hbox{\vrule height 8.pt depth 3.5pt width\z@}
\def\vfootnote#1{\insert\footins\bgroup
\baselineskip10pt\footfont
\interlinepenalty=\interfootnotelinepenalty
\floatingpenalty=20000
\splittopskip=\ht\strutbox \boxmaxdepth=\dp\strutbox
\leftskip=24pt \rightskip=\z@skip
\parindent=12pt \parfillskip=0pt plus 1fil
\spaceskip=\z@skip \xspaceskip=\z@skip
\Textindent{$#1$}\footstrut\futurelet\next\fo@t}
\def\Textindent#1{\noindent\llap{#1\enspace}\ignorespaces}
\def\footnote#1{\attach{#1}\vfootnote{#1}}%

\def\foot{\attach\footsymbolgen\vfootnote{\footsymbol}}
\let\footsymbol=\star
\newcount\lastf@@t           \lastf@@t=-1
\newcount\footsymbolcount    \footsymbolcount=0
\def\footsymbolgen{\relax\footsym
\global\lastf@@t=\pageno\footsymbol}
\def\footsym{\ifnum\footsymbolcount<0
\global\footsymbolcount=0\fi
{\iffrontpage \else \advance\lastf@@t by 1 \fi
\ifnum\lastf@@t<\pageno \global\footsymbolcount=0
\else \global\advance\footsymbolcount by 1 \fi }
\ifcase\footsymbolcount \fd@f\star\or
\fd@f\dagger\or \fd@f\ast\or
\fd@f\ddagger\or \fd@f\natural\or
\fd@f\diamond\or \fd@f\bullet\or
\fd@f\nabla\else \fd@f\dagger
\global\footsymbolcount=0 \fi }
\def\fd@f#1{\xdef\footsymbol{#1}}
\def\space@ver#1{\let\@sf=\empty \ifmmode #1\else \ifhmode
\edef\@sf{\spacefactor=\the\spacefactor}
\unskip${}#1$\relax\fi\fi}
\def\attach#1{\space@ver{\strut^{\mkern 2mu #1} }\@sf\ }
%
\newif\ifnref
\def\rrr#1#2{\relax\ifnref\nref#1{#2}\else\ref#1{#2}\fi}

\def\nrf#1{\nreftrue{#1}\nreffalse}

\def\multref#1#2#3{\nrf{#1#2#3}\refs{#1{--}#3}}
\nreffalse
\def\refout{\listrefs}
%
\def\eqn#1{\xdef #1{(\secsym\the\meqno)}
\writedef{#1\leftbracket#1}%
\global\advance\meqno by1\eqno#1\eqlabeL#1}
\def\eqnalign#1{\xdef #1{(\secsym\the\meqno)}
\writedef{#1\leftbracket#1}%
\global\advance\meqno by1#1\eqlabeL{#1}}
%
\def\chap#1{\newsec{#1}}
\def\chapter#1{\chap{#1}}
\def\sect#1{\subsec{#1}}
\def\section#1{\sect{#1}}
\def\\{\ifnum\lastpenalty=-10000\relax
\else\hfil\penalty-10000\fi\ignorespaces}
\def\note#1{\leavevmode%
\edef\@@marginsf{\spacefactor=\the\spacefactor\relax}%
\ifdraft\strut\vadjust{%
\hbox to0pt{\hskip\hsize\hskip.05in%
\vbox to0pt{\vskip-\dp\strutbox%
\sevenrm\baselineskip=10pt plus 1pt minus 1pt%
\ifx\answ\bigans\hsize=.9in\else\hsize=.4in\fi%
\tolerance=5000 \hbadness=5000%
\leftskip=0pt \rightskip=0pt \everypar={}%
\raggedright\parskip=0pt \parindent=0pt%
\vskip-\ht\strutbox\noindent\strut#1\par%
\vss}\hss}}\fi\@@marginsf\kern-.01cm}
\def\titlepage{%
\frontpagetrue\nopagenumbers\abstractfont%
\hsize=\hstitle\rightline{\vbox{\baselineskip=10pt%
{\abstractfont\pubnum}}}\pageno=0}
\frontpagefalse
\def\pubnum{}
\def\pdate{\number\month/\number\yearltd}
\def\makefootline{\iffrontpage\vskip .27truein
\line{\the\footline}
\vskip -.1truein\line{\pdate\hfil}
\else\vskip.5cm\line{\hss \tenrm $-$ \folio\ $-$ \hss}\fi}
\def\title#1{\vskip .7truecm\titlestyle{\titleft #1}}
\def\titlestyle#1{\par\begingroup \interlinepenalty=9999
\leftskip=0.02\hsize plus 0.23\hsize minus 0.02\hsize
\rightskip=\leftskip \parfillskip=0pt
\hyphenpenalty=9000 \exhyphenpenalty=9000
\tolerance=9999 \pretolerance=9000
\spaceskip=0.333em \xspaceskip=0.5em
\noindent #1\par\endgroup }
\def\autskip{\ifx\answ\bigans\vskip.5truecm\else\vskip.1cm\fi}
\def\author#1{\vskip .7in \centerline{#1}}

\def\address#1{\ifx\answ\bigans\vskip.2truecm
\else\vskip.1cm\fi{\it \centerline{#1}}}
\def\abstract#1{\vskip .5in\vfil\centerline
{\bf Abstract}\penalty1000
{{\smallskip\ifx\answ\bigans\leftskip 2pc \rightskip 2pc
\else\leftskip 5pc \rightskip 5pc\fi
\noindent\abstractfont \baselineskip=12pt
{#1} \smallskip}}
\penalty-1000}
\def\endpage{\tenpoint\supereject\global\hsize=\hsbody%
\frontpagefalse\footline={\hss\tenrm\folio\hss}}
%
\def\CERN{\address{CERN, Geneva, Switzerland}}
\def\inbar{\vrule height1.5ex width.4pt depth0pt}
\def\IC{\relax\,\hbox{$\inbar\kern-.3em{\mss C}$}}
\def\IF{\relax{\rm I\kern-.18em F}}
\def\IH{\relax{\rm I\kern-.18em H}}
\def\II{\relax{\rm I\kern-.17em I}}
\def\IN{\relax{\rm I\kern-.18em N}}
\def\IP{\relax{\rm I\kern-.18em P}}
\def\IQ{\relax\,\hbox{$\inbar\kern-.3em{\rm Q}$}}
\def\IR{\relax{\rm I\kern-.18em R}}
\def\ZZ{\relax{\hbox{\mss Z\kern-.42em Z}}}
\def\nup#1({Nucl.\ Phys.\ $\us {B#1}$\ (}
\def\plt#1({Phys.\ Lett.\ $\us  {B#1}$\ (}
\def\plb#1({Phys.\ Lett.\ $\us  {#1B}$\ (}
\def\cmp#1({Comm.\ Math.\ Phys.\ $\us  {#1}$\ (}
\def\prp#1({Phys.\ Rep.\ $\us  {#1}$\ (}
\def\prl#1({Phys.\ Rev.\ Lett.\ $\us  {#1}$\ (}
\def\prv#1({Phys.\ Rev. $\us  {#1}$\ (}
\def\und#1({            $\us  {#1}$\ (}
\def\tit#1,{{\it #1},\ }
%

\def\bar{\overline}
\def\us#1{\bf{#1}}
\def\hat{\widehat}

\def\Coe#1.#2.{{#1\over #2}}

\def\coe#1.#2.{\relax{\textstyle {#1 \over #2}}\displaystyle}

\def\notin{\hbox{{$\in$}\kern-.51em\hbox{/}}}

\catcode`\@=12

\def\FLST{\rrr\FLST{S. Ferrara,
         D. L\"ust, A. Shapere and S. Theisen, \plt225 (1989) 363.}}

\def\BA{\rrr\BA{I. Bars and K. Sfetsos, preprints USC-92/HEP-B2,
USC-92/HEP-B3.}}

\def\TSEY{\rrr\TSEY{A.A. Tseytlin, preprints Imperial/TP/92-93/7,
CERN-TH.6804/93.}}

\def\HET{\rrr\HET
{D.J. Gross, J.A. Harvey, E. Martinec and R. Rohm, \nup256 (1985)
253.}}

\def\PETE{\rrr\PETE{
M.J. Perry and E. Teo, preprint DAMTP R93/1; P. Yi, preprint
CALT-68-1852.}}

\def\IL{\rrr\IL{L. Ib\'a\~nez and D. L\"ust, \nup382 (1992) 305.}}

\def\KOUN{\rrr\KOUN{C. Kounnas, preprints CERN-TH.6790/93,
CERN-TH.6799/93.}}

\def\ALOS{\rrr\ALOS{E. Alvarez and M. Osorio, \prv40 (1989) 1150.}}

\def\KL{\rrr\KL{C. Kounnas and D. L\"ust, \plt289 (1992) 56.}}

\def\POL{\rrr\POL{J. Polchinski, \nup324 (1989) 123.}}

\def\BAKI{\rrr\BAKI{I. Bakas and E. Kiritsis, Int. J. Mod. Phys.
{\bf A7} (1992) 55; J. Ellis, N. Mavromatos and D.V. Nanopoulos,
\plt272 (1991) 261, \plt284 (1992) 43.}}

\def\DUAL{\rrr\DUAL
{K. Kikkawa and M. Yamasaki, \plt149 (1984) 357;
      N. Sakai and I. Senda, Progr. Theor. Phys. {\bf 75} (1986) 692.}}

\def\ROVE{\rrr\ROVE{M. Rocek and E. Verline, \nup373 (1992) 630;
A. Giveon and M. Rocek, \nup380 (1992) 128.}}

\def\WIT{\rrr\WIT{E. Witten, \prv44 (1991) 314.}}

\def\WZW{\rrr\WZW{E. Witten, Comm. Math. Phys. {\bf 92} (1984) 455;
E. Witten, \nup223 (1983) 422;
K. Bardacki, E. Rabinovici and B. Saering, \nup301 (1988) 151;
D. Karabali and H.J. Schnitzer, \nup 329 (1990) 649.}}

\def\DLP{\rrr\DLP{L. Dixon, J. Lykken and M. Peskin, \nup325 (1989)
325;
I. Bars, \nup334 (1990) 125; I. Bars and D. Nemeschansky, \nup348
(1991) 89.}}

\def\BUSCH{\rrr\BUSCH{T. Buscher, \plt194 (1987) 59, \plt201 (1988)
466;
L.E. Ib\'a\~nez, D. L\"ust, F. Quevedo and
S. Theisen, unpublished notes (1990);
E. Smith and J. Polchinski, \plt 263 (1991) 59;
G. Veneziano, \plt265 (1991) 287;
A.A. Tseytlin, Mod. Phys. Lett. {\bf A6} (1991) 1721;
X. de la Ossa and F. Quevedo, preprint NEIP-92-004.}}

\def\GIQU{\rrr\GIQU{P. Ginsparg and F. Quevedo, \nup385 (1992) 527.}}

\def\KIR{\rrr\KIR{E.B. Kiritsis, Mod. Phys. Lett. {\bf A6} (1991) 2871.}}

\def\KIRA{\rrr\KIRA{E. Kiritsis, preprint CERN-TH.6797/93;
A. Giveon and E. Kiritsis, preprint CERN-TH.6816/93.}}

\def\GIV{\rrr\GIV{A. Giveon, Mod. Phys. Lett. {\bf A6} (1991) 2843.}}

\def\RAB{\rrr\RAB{S. Elitzur, A. Forge and E. Rabinovici, \nup359
(1991) 581; G. Mandal, A.M. Sengupta and S.R. Wadia,
Mod. Phys. Lett. {\bf A6} (1991) 1685.}}

\def\DIVER{\rrr\DIVER{R. Dijgkraaf, E. Verlinde and H. Verlinde,
\nup371 (1992) 269.}}

\def\CFMP{\rrr\CFMP{C. Callan, D. Friedan, E. Martinec and M. Perry,
\nup262 (1985) 593.}}

\def\MUL{\rrr\MUL{M. Muller, \nup337 (1990) 37.}}

\def\HOR{\rrr\HOR{P. Horava, \plt278 (1992) 101.}}

\def\ANT{\rrr\ANT{I. Antoniadis, C. Bachas, J. Ellis and D.V. Nanopoulos,
\plt211 (1988) 393, \nup328 (1989) 117.}}

\def\MYE{\rrr\MYE{R.C. Myers, \plt199 (1987) 371;
K.A. Meissner and G. Veneziano, \plt267 (1991) 33;
A. Sen, \plt271 (1991) 295;
M. Gasperini, J. Maharana and G. Veneziano, \plt272 (1991) 277;
M. Gasperini and G. Veneziano, \plt277 (1992) 256;
A.A. Tseytlin, Class. Quant. Grav. {\bf 9} (1992) 979;
A.A. Tseytlin, preprint DAMPT-92-15;
A.A. Tseytlin, preprint DAMPT-92-36;
I. Bars  and K. Sfetsos, preprint USC-92/HEP-B1;
K. Behrndt, preprints DESY 92-055;
DESY 92-179;
M. Gasperini and G. Veneziano, preprint CERN-TH.6572/92;
H.J. de Vega and N. Sanchez, preprint LPTHE 92-31;
H.J. de Vega, A.V. Mikhailov and N. Sanchez, preprint LPTHE 92-32.}}

\def\MV{\rrr\MV{K.A. Meissner and G. Veneziano, Mod. Phys. Lett.
{\bf A6} (1991) 3397.}}

\def\TSVA{\rrr\TSVA{
R. Brandenberger and C. Vafa, \nup316 (1988) 391;
A.A. Tseytlin and
C. Vafa, \nup372 (1992) 443.}}

\def\NAWI{\rrr\NAWI{C. Nappi and E. Witten, \plt293 (1992) 309.}}

\def\HWANG{\rrr\HWANG{S. Hwang, \nup354 (1991) 100; H. Hennigson,
S. Hwang and P. Roberts, \plt267 (1991) 350.}}

\def\HAEL{\rrr\HAEL{
S.W. Hawking and G.F.R. Ellis, {\it ``The Large Scale Structure of
Space-Time'',} Cambridge University Press, Cambridge, 1973.}}

\def \ANFEKOU{\rrr\ANFEKOU{I. Antoniadis, S. Ferrara and C. Kounnas,
preprint LPTENS 91-30; I. Antoniadis and C. Kounnas, preprint
LPTENS 91-31.}}

\def\OFFSET{\hoffset=6.pt\voffset=40.pt}

\OFFSET

%


\def\pubnum{
\hbox{CERN-TH.6850/93}}

\def\pdate{March 1993}

\titlepage
\title
{Cosmological String Backgrounds}
\vskip-.8cm
\author{{\bf Dieter L\"ust}\foot{Heisenberg Fellow}}
\CERN
\bigskip
\vfil
{\centerline{\it Talk given at the }}
{\centerline{\it ``4th Hellenic School on Elementary Particle Physics"}}
{\centerline{\it Corfu, 2-20 September 1992}}
\bigskip
\bigskip\vfil
\abstract{The propagation of strings in cosmological space-time
backgrounds is reviewed. We show the relation of a special
class of cosmological backgrounds to exact conformal field theory.
Particular emphasis is put on the singularity structure
of the cosmological space-time and on the discrete duality
symmetries of the string background.}

\vskip2.5cm
\noindent CERN-TH.6850/93\endpage
\sequentialequations

\centerline{{ \ninebf COSMOLOGICAL STRING BACKGROUNDS}}
\vskip.8truecm
\centerline{{\ninerm Dieter L\"ust}}
\centerline{{\it CERN, CH 1211 Geneva 23, Switzerland}}
\vskip.5truecm
\vbox{\hbox{\centerline{{\ninerm ABSTRACT}}}
{\smallskip\leftskip 3pc \rightskip 3pc \noindent \ninerm
The propagation of strings in cosmological space-time
backgrounds is reviewed. We show the relation of a special
class of cosmological backgrounds to exact conformal field theory.
Particular emphasis is put on the singularity structure
of the cosmological space-time and on the discrete duality
symmetries of the string background.
\smallskip}}

\chap{Introduction}

String theory is at the moment
the most attractive candidate for a unified
description of the basic constituents in nature and their interactions.
In the past years, the main emphasis within this subject was
put on the discussion of critical strings: $D=10$ for the most
promising case of the heterotic string \HET. In order to get contact
with low energy phenomenology, one can make the consistent
assumption that the string background space has the form of
four-dimensional flat Minkowski space-time times a six-dimensional
internal, compact
space with characteristic size of order of the Planck scale.
More generally, four-dimensional heterotic strings are based on the
tensor product of  a trivial (super) conformal
field theory having $c_L=4$ ($c_R=6$), corresponding to flat
four-dimensional space time,
with an internal (super)
conformal field theory possessing $c_L=22$ ($c_R=9$).
The form of the internal compact space or internal conformal field
theory determines all particle physics properties of
the four-dimensional string, such as the  gauge group, the number of
supersymmetries (gravitinos), matter representations, Yukawa
couplings etc.

In order to investigate quantum gravity effects like processes
shortly after the `Big Bang', a further step is necessary:
the flat four-dimensional Minkowski space-time has to be
replaced by a curved target space. Consequently,
this part of the string theory  also corresponds to a non-trivial
conformal field theory.
In general relativity, singularities in curved space-times are often
unavoidable. In fact, the well-known singularity theorems \HAEL\
proof the existence of singularities under certain assumptions
of the matter energy momentum tensor. For example, the standard
Big Bang scenario exhibits an initial singularity
at $t=0$.
In string theory, there are several reasons to believe that
singularities in the target space are not harmful to the theory.
Heuristically, this believe is based on the fact that the string
theory possesses a `minimal length scale' set by the
extension of the string itself. A well known example of a finite
string theory with a singular background space is the compactification
of strings on orbifolds.

One of the keys in understanding the meaning of singularities
and of the minimal length scale in string
theory may be given by the socalled duality symmetries \DUAL.
Duality is, roughly spoken, the invariance
of the string theory under the inversion of certain characteristic
length parameters. More generally, duality transformations
act on the background parameters of the string, like the space-time
metric or dilaton field. The existence of the duality
symmetry was first shown in the context of string compactification
on a constant background, like a circle or torus.
Here the stringy nature of this symmetry can be clearly identified
since duality involves the exchange of momentum with stringy
winding modes. The emergence of the duality symmetries
within string compactifications already proved to be very
useful \FLST\ to get information about the effective Lagrangian
in `flat' four-dimensional string theories.
For example one can show \IL\
the certain orbifold compactifications cannot lead to the spectrum
of the minimal supersymmetric standard model without breaking
the stringy duality symmetries.

The duality symmetries are not limited  to flat backgrounds;
their existence was shown \BUSCH,\KIR,\GIQU\ for many curved, for example
time-dependent backgrounds.
Even for non-compact spaces, duality survives as
discussed recently in ref.\KIRA.
This observation is of particular interest in the context
of discussing singularities in curved string backgrounds.
For non-compact spaces, the duality transformations
can change the topology of the space. This topology
change is most dramatic when the duality transformation
maps a singular on a non-singular  space, as shown first
for the Euclidean two-dimensional black-hole in ref.\GIV.
Since both spaces are equivalent from the string point of view,
the meaning of the singularity in the target space is unclear.

In this review we will discuss the propagation of strings
in cosmological, i.e. time-dependent
backgrounds \multref\MYE{\ANT\TSVA\MUL\MV\KL}\NAWI.
In the next section we will investigate the time-dependent
solutions of the string equations of motion including a non-trivial
metric and dilaton background. In section 3 we will
relate particular cosmological string backgrounds
to exact conformal field theories, namely the gauged
Wess-Zumino-Witten (WZW) models. Finally, in section four,
we will discuss the cosmological background in the socalled Einstein
frame. Particular emphasis is put on the duality symmetries.

\chap{Cosmological String Backgrounds}

Let us consider the propagation of a (bosonic) string
in the presence of a $D$-dimensional
metric and dilaton background
$G_{MN}(x)$ ($M,N=0,\dots,D-1$) and $\Phi(x)$.
The string tree level
effective action for these background fields has the form
$$S_{\rm eff}=\int{\rm d}^Dx\sqrt{-G}e^\Phi(R -(D\Phi)^2
+\Lambda).\eqn\effeaction
$$
The cosmological constant is related to the dimension $D$
of space time as $\Lambda={2(26-D)\over 3}$.
The effective action \effeaction\
leads to the following equations of motion
$$\eqalign{&R_{MN}+D_MD_N\Phi=0,\cr
&R+(D_M\Phi)^2+2D_MD^M\Phi=\Lambda.\cr}\eqn\sigmaeq
$$
These equations can be also obtained  \CFMP\
as the conditions of vanishing $\beta$-functions in the
corresponding two-dimensional $\sigma$-model
$$S_{2-{\rm dim}}=\int{\rm d}^2\sigma\sqrt g\lbrack g^{mn}G_{MN}(x)
\partial_m X^M\partial_n X^N-{r^{(2)}\over 4}\Phi(x) \rbrack
.\eqn\sigmaaction
$$
As an ansatz for the metric and dilaton field
we consider a cosmological, i.e. time dependent background of the form
$$G_{00}=-1,\qquad G_{ij}=\delta_{ij}R_i(t)^2\quad(i,j=1,\dots, D-1),
\qquad\Phi=\Phi(t).\eqn\ansatz
$$
Then the string field equations \sigmaeq\
can be rewritten as \MUL
$$\eqalign{&\sum_i{\ddot R_i\over R_i}+\ddot\Phi=0,\cr
&
{\ddot R_i\over R_i}+\sum_{j\neq i}{\dot R_i\over R_i}{\dot R_j\over R_j}
+\dot\Phi{\dot R_i\over R_i}=0,\cr
&\ddot\Phi+{1\over 2}
\Lambda+{1\over 2}\dot\Phi^2-\sum_{i<j}{\dot R_i\over R_i}
{\dot R_j\over R_j}=0.\cr}\eqn\timedepeq
$$

Now, it is not difficult to realize that the field equations \timedepeq\
are invariant under the duality transformation
$$\eqalign{R_i(t)&\rightarrow {1\over R_i(t)},\cr
\Phi(t)&\rightarrow\Phi(t)+2\log R_i(t).\cr}\eqn\dualtr
$$
The non-trivial duality transformation behavior of the
dilaton field implies that the time-dependent string coupling
constant is transformed like $g^2(t)=e^{-\phi(t)}\rightarrow g^2(t)
R_i(t)^{-2}$. This change of the string coupling constant agrees
with the transformation  of $g^2$ in the static case when one considers
the genus expansion of the string partition function \ALOS.
The origin of the duality invariance comes from the fact that the
metric \ansatz\ is independent from the spatial coordinates $x_i$
which leads to $D-1$ Abelian isometries of the model. In this case it
follows that the non-linear $\sigma$-model based on the dual
background \dualtr\ corresponds to a conformal field theory provided
that also the original background is conformal (at lowest order
in $\alpha'$) \ROVE.

The explicit form of the
solutions \MUL\ of the field equations \timedepeq\ depends on the
value of the cosmological constant $\Lambda$.
For $\Lambda=0$ ($D=26$) one obtains a family of solutions
$$\eqalign{&R_i(t)=\alpha_i(t-t_0)^{p_i},\qquad\sum_{i=1}^{D-1}p_i^2=1
\cr &e^\Phi= \beta^2( t-t_0)^p,\qquad p=1-\sum_{i=1}^{D-1}p_i,\cr}
\eqn\soluta
$$
where $\alpha_i,\beta,t_0$ are arbitrary real parameters.

For $\Lambda\neq 0$, the cosmological solutions are given by
$$\eqalign{&R_i(t)=\alpha_i\lbrack{\rm tanh}{\sqrt{-\Lambda}\over 2}
(t-t_0)\rbrack^{p_i},\qquad\sum_{i=1}^{D-1}p_i^2=1
\cr &e^\Phi= \beta^2
\lbrack{\rm cosh}{\sqrt{-\Lambda}\over 2}
( t-t_0)\rbrack^{2-p}
\lbrack{\rm sinh}{\sqrt{-\Lambda}\over 2}
( t-t_0)\rbrack^{p}
,\qquad p=1-\sum_{i=1}^{D-1}p_i
.\cr}
\eqn\solutb
$$
Alternatively, for $\Lambda\neq 0$, there is another solution,
the socalled linear dilaton \ANT:
$$R_i={\rm const},\qquad \Phi=\sqrt{-\Lambda}t.\eqn\solutc
$$

To understand the singularity structure and the duality properties of
the above cosmological solutions let us compute the Ricci-tensor.
For $\Lambda=0$ it has the form ($\alpha_i=1,t_0=0$):
$$R_{00}={p\over t^2},\qquad
R_{ij}=\delta_{ij}{p_ip\over t^{2-2p_i}}.\eqn
\ricci
$$
The corresponding scalar curvature is then
$$R=-{p^2\over t^2}.\eqn\scalarc
$$
We see that for arbitrary $p$, space-time is singular at $t=0$.
However for $p=0$, space-time is Ricci-flat. However the curvature
tensor is still singular at $t=0$ for this case ($R_{ab}^{cd}\sim t^{-2}
$). Only for the ``two-dimensional'' case $p_1=1$, $p_i=0$ ($i\neq 1$)
space-time is completely flat. This is most easily seen by introducing
lightcone coordinates like $u=-te^{-x_1}$, $v=te^{x_1}$.
Then the metric becomes that of flat Minkowski space time,
${\rm d}s^2={\rm d}u{\rm d}v$, where the whole
$t-x_1$ plane is mapped onto the forward/backward lightcones $uv<0$.

For $\Lambda\neq 0$, the scalar curvature has the form
$$R=\Lambda{(4-4p)({\rm sinh}{\sqrt{-\Lambda}\over 2}t)^2
-p^2\over ({\rm sinh}{\sqrt{-\Lambda}\over 2}t)^2
({\rm cosh}t{\sqrt{-\Lambda}\over 2})^2}.
\eqn\scalarb
$$
Thus for generic values of $p$ there is a singularity at $t=0$.
However for $p=0$, $\Lambda<0$, the scalar curvature is finite for all
$t$ \MV\ whereas
the curvature tensor is still singular at $t=0$. Only in
the ``two-dimensional'' case with $p_1=1$ and $p_i=0$ ($i\neq 1$)
space-time is completely free of singularities in the $t$-$x_1$ plane.

The duality transformation $p_i\rightarrow -p_i$, $p\rightarrow p+2p_i$
relates two inequivalent cosmological
backgrounds. For generic $p$, the original space-time and also
the dual transformed space-time possess a singular scalar
curvature. However for $p=0$,
a space-time with singular scalar curvature is mapped onto a
background without singularity in $R$. In particular, for $p_1=1$
and $\Lambda=0$, flat Minkowski space-time is mapped onto
a singular space, and for $p_1=1$, $\Lambda< 0$, a singular-free
expanding space is mapped onto a singular, but contracting
space-time. (For more discussion on this case see next chapter.)

The equations \sigmaeq\ and \timedepeq\ are the tree-level
field equations determining the string propagation in the
graviton, dilaton vacuum background. However to obtain
a realistic cosmological scenario one has to include also
matter energy density and pressure from the stringy matter.
As discussed in ref.\TSVA\ this is determined by the one-loop
free energy of the string. In an isotropic universe with $R_l(t)
=R(t)$ ($l=1,\dots, d$, all other $R_i={\rm constant}$), $\Lambda=0$,
assuming that the matter evolves
adiabatically, one can show \TSVA\ that the
massless momentum modes with energy $E\propto R(t)^{-1}$
lead to the radiation dominated era of the standard cosmology with
$R(t)\propto e^{{2\over d+1}}$, and with a dilaton approaching a
constant.
However at very early times the contribution of the winding
modes to the energy also becomes important. These modes oppose
the expansion of the universe, and, as it was argued in \TSVA,
the universe oscillates for some time around the Planck scale,
until by coincidence
it starts expanding in a smaller number
of dimensions ending in the radiation dominated era.
Based on the thermodynamics of the winding modes one can give
arguments \TSVA\ that the preferred number of expanding directions is
smaller than four.

\chap{Gauged WZW Models}

Now we want to relate the cosmological solutions of the
string field equations to exact conformal field theories, namely
to the gauged WZW models. We will show \KL\ that the ``two-dimensional''
model with $p_1=\pm 1$ can be derived from
a gauged Wess-Zumino-Witten (WZW) models \WZW\ based on the
non-compact coset space $SL(2,{\bf R})/SO(1,1)$.
In fact it was first shown by Witten \WIT\ that this type
of coset space CFT has a very interesting target space interpretation,
namely that of a two-dimensional black-hole \RAB.
However a simple change
in the CFT, namely the change of sign of the level $k$ of the
$SL(2,{\bf R})$ Kac-Moody algebra, will lead to the cosmological
interpretation.

The gauged WZW model based on the coset $G/H$ is described by the
action
$$\eqalign{S=&{k\over 4\pi}\int {\rm d}^2z{\rm tr}(g^{-1}\partial \
g^{-1}\bar\partial g)-{k\over 12\pi}\int_B{\rm tr}(g^{-1}dg\wedge g^{-1}
dg\wedge g^{-1}dg)\cr
&+{k\over2\pi}\int{\rm d}^2{\rm tr}(A\bar\partial gg^{-1}-\bar Ag^{-1}
\partial g-g^{-1}Ag\bar A),\cr}\eqn\wzwaction
$$
where the boundary of $B$ is the 2D worldsheet,
$g$ is a group element of the group $G$, and $A$ are the gauge fields
of $H$ transforming as $A\rightarrow h_L^{-1}(A+\partial )h_L$; $k$ is
the level of the Kac--Moody algebra for the group $G$. To be specific,
we want to discuss a four-dimensional target-space
based on the non-compact coset $SL(2,{\bf R})\times SO(1,1)^2/SO(1,1)$.
The central charge of this coset CFT (non-compact coset CFT's were
discussed in \DLP)
is given by
$$c=4+{6\over k-2},\eqn\central
$$
where the $k$, being a real number, is the level of the non-compact
$SL(2,{\bf R})$ Kac--Moody algebra.
For the case that the string theory
is entirely given by this coset CFT, the condition $c=26$ implies
$k=25/11$.
However, we leave $k$ as a free parameter. Therefore we
couple the $SL(2,{\bf R})\times SO(1,1)^2/
SO(1,1)$ coset CFT to an internal CFT with central charge
$$c_{\rm int}=D-4=
22-\delta c,\qquad
\delta c=26-D={6\over k-2}.\eqn\centralch
$$

Now we parametrize the
group element of $SL(2,{\bf R})$ as
$g=\pmatrix{u&a\cr -b&v\cr}$ with $uv+ab=1$.
Finally we assume for simplicity that $H=SO(1,1)$
is entirely inside $SL(2,{\bf R})$. Performing a vector-like
gauging, $a=\pm b$, the action \wzwaction\ can be identified with
a $\sigma$-model action of the form
$$S=\int{\rm d}^2zG_{MN}
\partial X^M\bar\partial X^N.\eqn\sigmaaction
$$
In the semiclassical approximation $k\rightarrow\pm\infty$,
the corresponding four-dimensional $\sigma$-model metric is given as
\WIT,\GIQU
$${\rm d}s^2=-k{{\rm d}u{\rm d}v \over 1-uv}+{\rm d}x_2^2
+{\rm d}x_3^2.\eqn\sigmametric
$$
The associated dilaton can be derived from the change of the
integration measure in the path integral. It has the form
$$\Phi=\log(1-uv)\eqn\dilaton
$$

We recognize that the signature of the two-dimensional part of
this metric depends just on the sign of the level $k$ of the
$SL(2,{\bf R})$ Kac--Moody algebra. In fact, for $k\rightarrow +\infty$
space-time possesses a singularity in futures times $\tau=u+v$.
In this
case, $u$ and $v$  are Kruskal-like coordinates of a black-hole metric;
more precisely the metric \sigmametric\ has the causal structure
of a four-dimensional black-brane (see figure 1).
The singularity at $uv=1$ originates from the fixed points
of the modded vector gauge symmetry $H$ at this curve.
However for $k\rightarrow -\infty$
the causal structure is completely different \KL. Specifically the
causal structure for negative $k$ is obtained from the black-hole case
by a $90^0$ rotation of the two-dimensional $u,v$-plane (see figure 2).
Specifically, we see
that,
introducing the proper time $\tau=u-v$, there is  no singularity
for future times $\tau$ inside the light-cone $uv<0$
(region I in figure
2). The singularity at $uv=1$ is hidden behind the horizon $uv=0$.
However signals in
regions II, III
may hit the singularity, and the singularity may also send signals
through regions II, III into region I.
Now one has  a forward light-cone with the
singularity behind it.
As we will discuss in the following, this class
of string backgrounds with negative $k$
describes an expanding Universe with singularity
outside the visible horizon.\foot{In ref.\TSVA\ it was discussed
that for $k>0$, $\Lambda>0$,
the interior region II of the two-dimensional
black-hole has an interesting cosmological interpretation, in
particular after going to an infinite cover fold of $SL(2,{\bf R})$.}

Using eq.\centralch\ is easy to determine for which number of
dimensions, i.e. for what values of central charges $\delta c$,
one has a black-hole or cosmological metric respectively.
Specifically for $D<26$ we deal with a black-hole.
(In addition, for $D<29$ one obtains again a black-hole metric.
Then, however, the central charge of $SL(2,{\bf R})/SO(1,1)$ becomes
negative, and one may expect serious problems with unitarity.)
On the other hand,
for $26<D<29$ one deals with the cosmological scenario. Regarding
this theory as a non-critical Liouville string in the asymptotic limit,
it would imply that the Liouville field is a time-like coordinate \POL.

To make contact with the discussion of section 2,
it is easy to show that inside the singularity free
region I we have in fact an expanding Universe.
Specifically, we
introduce coordinates which cover exactly the region I,
$$u=e^{x_1/\sqrt{-k}
}{\rm sinh}({t\over\sqrt{-k}})
,\qquad v=-e^{-x_1/\sqrt{-k}}{\rm sinh}({t\over\sqrt{-k}})
.\eqn\coordtwo
$$
(The coordinates $t$ and $x_1$ are not geodesically complete.)
Then the metric \sigmametric\ in region I becomes
$${\rm d}s^2=-{\rm d}t^2+\lbrack{\rm tanh}({t\over\sqrt{-k}})
\rbrack^2{\rm d}x_1^2
+{\rm d}x_2^2+{\rm d}x_3^2.\eqn\metrictwo
$$
This metric is exactly of the form eq.\solutb\ with $p=0$, $p_1=1$,
all other $p_i=0$ and $\Lambda={2\delta c\over 3}={4\over k-2}$
after the necessary
shift $k\rightarrow k-2$ which originates from renormalization
effects.

In the gauged WZW model, the
duality operation corresponds to the exchange of
gauging the axial $SO(1,1)$ subgroup of $SL(2,{\bf R})$
instead of gauging the vector-like subgroup $SO(1,1)$ \KIR,\DIVER.
This implies that in the $u,v$ coordinate system, duality acts as
$$uv\rightarrow uv-1.\eqn\dualuv
$$
This is just the exchange of the regions
I and IV in figures 1 and 2, whereas the regions II and III
are mapped onto
themselves \GIV,\DIVER.
In the $t,x_1$ coordinate system, as defined in eq.\coordtwo,
the duality transformation \dualtr\ leads to the dual metric of the
form
$${\rm d}s^2_D=-{\rm d}t^2+\lbrack{\rm coth}({t\over\sqrt{-k}})
\rbrack^2{\rm d}x_1^2
+{\rm d}x_2^2+{\rm d}x_3^2.\eqn\metrictwodual
$$
and the dual dilaton
$$\Phi(t)_D=
2\log{\rm sinh}({t\over\sqrt{-k}}).\eqn\diltrans
$$
Therefore the dual metric of region IV possesses a singularity
at $t=0$.

The coset CFT based on $SL(2,{\bf R})/SO(1,1)$ with negative
level $k=-N$ is closely related to the CFT of the coset space
$SU(2)/U(1)$ of level $N$ with the same central
charge $c={3N\over N+2}-1$.
The gauged WZW model of $SU(2)/U(1)$ leads, for large $N$, to
a $\sigma$-model with the following
two-dimensional target space metric:
$${\rm d}s^2=
N{{\rm d}z{\rm d}\bar z\over
1-z\bar z }
.\eqn\sutwoa
$$
Here $z$ is a complex parameter of an $SU(2)$ group element,
$z=x+i\tau$.
Performing a change of coordinates like
$$z=e^{ix_1/\sqrt{N}
}\sin({x_0\over\sqrt{N}})
,\eqn\euclidcoor
$$
the metric \sutwoa\ becomes
$${\rm d}s^2={\rm d}x_0^2+\lbrack\tan({x_0\over\sqrt{N}})
\rbrack^2{\rm d}x_1^2.\eqn\euclidmetr
$$
This Euclidean
metric is just the analytic continuation of the cosmological
metric \metrictwo, i.e $x_0=-it$.
It leads to a singularity at $x_0=\pi /2$.
In contrast to its Minkowski counter part, the Euclidean
space is self-dual, which means that the duality transformation
\dualtr\ just acts as a `time' reparametrization $x_0\rightarrow
x_0+\pi/2$.
It is interesting to note that conformal field theory $SL(2,{\bf R})/
U(1)$ with positive $k$, which leads to the two-dimensional black-hole,
is governed by an infinite dimensional, non-linear symmetry
$\hat W_\infty(k)$ \BAKI. This algebra truncates for the
cosmological case $k=-N<0$ to the finite algebra $W_N$. Both algebras
linearize in the limit $k,N\rightarrow \infty$, where the ordinary
$W_\infty$ algebra is recovered.

So far we have considered the semiclassical approximation
of the gauged WZW model, i.e. the target space metric, as given
in eq.\sigmametric, was derived for $k\rightarrow\pm\infty$.
This metric, together with the dilaton eq.\dilaton\ provides
a solution of the string field equations eq.\sigmaeq, which are
valid at lowest order in $\alpha'$ or $1/k$.
Thus in the bosonic $\sigma$-model there are higher order
corrections to the metric and dilaton background. However
for models with extended (n=4) worldsheet supersymmetries,
there are no renormalization effects, and the lowest order
background is supposed to be exact (see the discussion
at the end of this chapter).
For the bosonic case, the exact,
i.e. conformal, background can be obtained by various
methods \DIVER,\BA,\TSEY.
For the coset $SL(2,{\bf R})/SO(1,1)\times{\bf R}^2$
the exact, all order metric has the following form \BA:
$${\rm d}s^2=-{k-2\over 1-(1-{2\over k})uv}\biggl\lbrack
{\rm d}u{\rm d}v +{1\over 2k(1-uv)}(v{\rm d}u+u{\rm d}v)^2\biggr\rbrack
+{\rm d}x_2^2
+{\rm d}x_3^2.\eqn\sigmametricexact
$$
Here $k-2$ is the renormalized coupling constant. Obviously,
this metric approaches the leading expression eq.\sigmametric\
for $k\rightarrow\pm\infty$. The exact dilaton has the form
$$e^\Phi=(1-uv)\sqrt{1+{2uv\over k(1-uv)}}.\eqn\dilatonexact
$$

To discuss the cosmological (or black-hole)
interpretation of the exact metric,
it is useful to again change the coordinates. For example, for
the region IV, $uv>1$, the new coordinates are given as
$$u=e^{x_1/\sqrt{-k+2}
}{\rm cosh}({t\over\sqrt{-k+2}})
,\qquad v=e^{-x_1/\sqrt{-k+2}}{\rm cosh}({t\over\sqrt{-k+2}})
.\eqn\coordtwo
$$
Then eq.\sigmametricexact\ becomes \DIVER,\BA
$${\rm d}s^2=-{\rm d}t^2+{1\over
\lbrack{\rm tanh}({t\over\sqrt{-k+2}})
\rbrack^2-{2\over k}}{\rm d}x_1^2
+{\rm d}x_2^2+{\rm d}x_3^2.\eqn\metricexact
$$
Examining this expression one observes the interesting fact
that for the cosmological case, $k<0$, there is no singularity in
space-time anymore, since
$1/(1+{2\over N})<R_1(t)^2_{\rm Exact}=\biggl(
\lbrack{\rm tanh}({t\over\sqrt{-k+2}})
\rbrack^2-{2\over k}\biggr)^{-1}<N/2$ for all $t$ ($N=-k>2$).
(However in the $n=4$ supersymmetric case the singularity is still
present.)
This is quite different from the lowest order metric
eq.\metrictwodual.\foot{For the black-hole case with $k>2$ it was
shown \PETE\ that the exact metric is also free of singularities in the
sense that the singularity is separated from the horizon by a
Euclidean region.} Thus in the exact CFT,
the physical relevance of the singularity
of the lowest order solution seems to be rather low. This observation
is also supported by the expectation that all physical amplitudes in
the coset CFT will be non-singular.

The duality transformation for the exact metric is again given
by $uv\rightarrow uv-1$ due to the exchange of vector versus
axial gauging. This implies that the exact dual metric is obtained
from eq.\metricexact\ by replacing
${\rm tanh}({t\over\sqrt{-k+2}})$ by
${\rm coth}({t\over\sqrt{-k+2}})$. However note that this transformation
does not anymore correspond to $R_1(t)_{\rm Exact}\rightarrow 1/
R_1(t)_{\rm Exact}$.

At the end of this section
let us discuss briefly some extended four-dimensional models.
For example consider
the gauged WZW model based on the
coset ${SL(2,{\bf R})\over SO(1,1)}
\times {SU(2)\over U(1)}$ \HOR,\KL,\NAWI,\KOUN.
The central
charge of this CFT is given by
$$c={3k_1\over k_1-2}+{3k_2\over k_2+2}-2.\eqn\centralcal
$$
The corresponding
background metric has the following form:
$${\rm d}s^2=-k_1{1\over 1-uv}
{\rm d}u{\rm d}v+k_2{1\over 1-z\bar z}{\rm d}z{\rm d}\bar z.\eqn\sutwo
$$
For the black-hole case with positive $k_1$
one can set
$k_1=k_2+4$. Then the model possesses an extended $n=4$ world-sheet
supersymmetry \KOUN, and
one obtains that $c=4$ regardless of the value of
$k_1$.
Thus $\delta c=0$ for all $k_1$
and the metric \sutwo\
is valid for all $k_1$, since there are no renormalization effects
which can lead to higher order contributions in $1/k$.

For the cosmological case with negative $k_1$ we cannot set $k_1=k_2+4$
($k_2$ is always positive).
As an alternative one can couple $SL(2,{\bf R})/SO(1,1)$
with negative $k_1$ to an Euclidean two-dimensional
`black-hole' based on the coset $SL(2,{\bf R})/U(1)$ with positive
Kac--Moody level $k_2$.
The central charge of this model is
$$c={3k_1\over k_1-2}+{3k_2\over k_2-2}-2.\eqn\centralcala
$$
and the background metric looks like
$${\rm d}s^2=-k_1{1\over 1-uv}
{\rm d}u{\rm d}v+k_2{1\over 1+z\bar z}{\rm d}z{\rm d}\bar z.\eqn\sutwobh
$$
Again one
obtains $\delta c=0$ for $k_1=-k_2+4$, where the model is $n=4$
supersymmetric.

\chap{The Einstein Frame}

To get contact with standard gravity theory, which possesses
a canonical Einstein term, one has to perform a Weyl rescaling
of the $\sigma$-model metric eq.\sigmametric\ by the exponential
of the dilaton field.\foot{There is no Weyl rescaling of the metric
in two space-time dimensions.}
This Weyl rescaling however does not change the
causal structure of the theory.
Specifically, the metric in the Einstein
frame is given as
$${\rm d}s^2=e^\Phi{\rm d}s^2_\sigma={\rm d}u{\rm d}v
+(1-uv)({\rm d}x_2^2+{\rm d}x_3^2).\eqn\einsteinmetric
$$
Here we have focused on the
cosmological case with $k\rightarrow -\infty$,
where we have absorbed the level $k$ in the metric. The effective
action now has the form
$$S_{\rm eff}=\int{\rm d}^4x\sqrt G\biggl({1\over 2}R-{1\over 4}D_M\Phi
D^M\Phi-V(\Phi)\biggr) ,\qquad V(\Phi)=2e^{-\Phi}.\eqn\einsteinaction
$$

To show the cosmological behavior of the metric inside
region I we again introduce
coordinates which cover exactly the region I,
namely
$$u=e^{x_1}t,\qquad v=-e^{-x_1}t.\eqn\change
$$
It follows that curves of constant times $t$ correspond in figure 2 to
hyperbolae $uv=-t^2$, whereas the curves of constant $x_1$ are given
by the straight lines $u/v=-e^{2x_1}$ (see figure 2).
In these coordinates
the metric now looks like \KL
$${\rm d}s^2=-{\rm d}t^2+t^2{\rm d}x_1^2
+(1+t^2)({\rm d}x_2^2+{\rm d}x_3^2),\eqn\newmetric
$$
and the dilaton has the form
$$\Phi(t)=\log(1+t^2).\eqn\newdil
$$
Clearly, the metric \newmetric\ describes an expanding Universe in region
I with two different scale factors $R_1(t)=t$, $R_{2,3}(t)=\sqrt{1+t^2}$,
where $t$ is the cosmological time coordinate.
For small $t$, the Universe expands unisotropically. However, for
large times, one approaches an isotropic, linear expansion of the
Friedmann-Robertson-Walker type with $R_i(t)=t$ ($i=1,2,3$).
However our solution has no initial singularity at $t=0$ in contrast to
the standard isotropic Robertson-Walker Universe.
(The Ricci tensor in the coordinates \change\ takes
the form $R_{tt}={2\over (1+t^2)^2}$, $R^i_j=-{2\over 1+t^2}\delta^i_j$.)

Let us derive the energy momentum tensor of the dilaton matter
field.
Specifically consider the
classical Einstein equations
$$R_{MN}-{1\over 2}G_{MN}R=-T_{MN}.\eqn\einsteinequ
$$
The corresponding
energy-momentum tensor from the dilaton matter field has
the form
$$T_{MN}={1\over 2}D_M\Phi D_N\Phi-G_{MN}\biggl({1\over 4}G^{PQ}D_P\Phi
D_Q\Phi+V(\Phi)\biggr) .\eqn\energymom
$$
Then we obtain, with $\Phi=\log(1+t^2)$ and $V(\Phi)={2\over 1+t^2}$,
that
$$\eqalign{
T_{tt}&={(\partial_t\Phi)^2\over 4}+V=\rho={3t^2+2\over (1+t^2)^2},\cr
T_{ij}&=\biggl({(\partial_t\Phi)^2\over 4}-V\biggr)G_{ij}
=pG_{ij}=-{t^2+2\over (1+t^2)^2}G_{ij}.\cr}\eqn\densitypres
$$
Here $\rho$ is the energy density of the dilaton matter system
and $p$ is its pressure. Now it is easy to see that the quantity
$$\rho+3p=-{4\over (1+t^2)^2}\eqn\rhopi
$$
is negative for all $t$.
It is interesting to observe that the form of $\rho+3p$, being always
negative, violates
an assumption by Hawking and Ellis \HAEL\
on the form
of the matter energy-momentum tensor, which, being satisfied,
would always lead to a singular space-time.
Thus, the absence of an initial
singularity in the cosmological region I can
be understood from the specific form of the energy-momentum tensor
of the dilaton matter system.

The duality transformation
in the Einstein frame
is expressed as
$$t^2\rightarrow -1-t^2.\eqn\dualt
$$
Thus we see again that the cosmological region I is mapped to
the cosmological region IV, which requires an analytic continuation
to imaginary $t$ values.
However it is more convenient to use real coordinates which cover
the region IV of the following form:
$$u=e^{x_1}t,\qquad v=e^{-x_1}t, \qquad t^2>1.\eqn\changea
$$
Then the dual metric can be written as
$${\rm d}s^2_D=-{\rm d}t^2+t^2{\rm d}x_1^2
+(t^2-1)({\rm d}x_2^2+{\rm d}x_3^2),\eqn\newmetricd
$$
and the duality transformed dilaton is obtained as
$$\Phi(t)_D=\log(t^2-1).\eqn\newdila
$$
It is easy to show that this metric leads to a scalar curvature which
is singular at $t=1$.
It is instructive to compute again the dual energy density and the dual
pressure of the dual dilaton matter system:
$$
\rho_D={3t^2-2\over (t^2-1)^2},\qquad
p_D={2-t^2\over (t^2-1)^2}.\eqn\densitypresd
$$
Thus the quantity $\rho_D+3p_D={4\over(t^2-1)^2}$ is now positive
which implies the existence of an initial singularity according to the
theorem of \HAEL.

\chap{Summary}

We have reviewed the cosmological solutions of the string
background equations and the action of the duality transformation
on the cosmological backgrounds.
In two dimensions (and possibly also in higher dimensions),
the cosmological coordinates  are not geodesically complete
and there a exist a Kruskal-like  completion of space-time.
The cosmological metric in the Kruskal coordinates
can be obtained from an exact CFT, the gauged WZW-models
based on the coset $SL(2,{\bf R})/SO(1,1)$.
The same CFT leads to the two-dimensional black-hole
when one changes the sign of the level of the underlying
non-compact Kac--Moody algebra. For both signs, black-holes as well as
cosmological backgrounds, the metric of the target space, at lowest
order in $\alpha'$,
leads to
space-time singularities. In the cosmological case the singularity is
hidden behind
the light-cone. However the singularity
could send signals into the light-cone and therefore influence
the expansion of the Universe. This is very similar to the initial
singularity (Big Bang) in the standard Friedmann-Robertson--Walker
Universe.
It is important to stress that, seen from the CFT
point of view, the singularities in the black-hole
as well as in the cosmological
frameworks have exactly the same origin.
Therefore
one could expect that the same type of quantum gravity effects
are relevant near the black-hole as well as in the early Universe
at times shortly
after the initial singularity.
These quantum gravity effects should in principle be determined from the
underlying coset CFT. (For considerations in this directions
for the black-hole case, see ref.\DIVER.)
Let us mention that the
elimination of the negative-norm
states for the gauged WZW model with negative level $k$ has still
to be demonstrated. However we believe that the unitarity of the spectrum
is possible since it was already shown \ANT\
that for the asymptotic region $t
\rightarrow\infty$, where one approaches the linear dilaton model,
the elimination of the negative norm states is possible.
In addition, the construction of a modular invariant partition function
is still an interesting problem.

\bigskip
\bigskip

I would like to thank C. Kounnas for a very pleasant collaboration
on many of the subjects presented here. I am also grateful
to E. Kiritsis for useful discussions.

\refout
\endpage

\centerline{\bf Figure Captions}
\vskip1cm

\noindent{\bf Figure 1:}
The causal structure of the two-dimensional slice of the black-hole
metric equation \sigmametric\ with positive Kac--Moody level $k$.

\noindent{\bf Figure 2:}
The causal structure of the two-dimensional slice of the cosmological
metric equation \sigmametric\ with negative Kac--Moody level $k$.

\vfill\eject\bye